\documentclass[12pt]{article}
\usepackage{amsmath,amssymb}
\usepackage{graphicx} 
\usepackage{epsfig}
\newcommand{\sect}[1]{\section{#1}\setcounter{equation}{0}}

\newcommand{\be}{\begin{equation}}
\newcommand{\ee}{\end{equation}}
\newcommand{\nn}{\nonumber}
\newcommand{\s}[1]{{\bf#1}}                  

\def\to{\rightarrow}
\def\half{\frac{1}{2}}
\def\ihalf{\tfrac{i}{2}}
\def\quart{{\frac{1}{ 4}}}

\def\tilde{\widetilde}
\def\tr{\hbox{Tr}}
 \def\eqalign#1{%
\null\,\vcenter{\openup\jot\m@th
  \ialign{\strut\hfil$\displaystyle{##}$&$\displaystyle{{}##}$\hfil
      \crcr#1\crcr}}\,}
\input{epsf.tex}
  
\begin{document}

\rightline{v2}

\bigskip
 \centerline{\bf A note on the spinor construction of Spin Foam amplitudes}
\bigskip
\centerline{ Giorgio Immirzi  }
 \medskip
\centerline{ Colle Ballone, Montopoli di Sabina (Italy)   }

\centerline{\footnotesize e-mail: giorgio.immirzi\@@pg.infn.it }

\medskip
\bigskip
\abstract{\footnotesize I discuss the use of spinors in the construction of spin-foam models,
in particular the form of the closure and simplicity constraints  for triangles that  are space-like,
i.e. with (area)$^2=\half S^{IJ}S_{IJ}>0$, regardless of whether  they belong
to  tetrahedra with  a space-like or time-like normal, emphasizing the role of the 
light-like 4-vector $u_t\sigma^I\bar u_t$.  
In the quantization of the model, with the representations of  SL(2,$\mathbb{C}$)
acting on spaces of functions of light-like vectors, one may use 
the canonical basis of SU(2) representations,  or the pseudobasis limited to the
discrete representations of SU(1,1); in alternative it is proposed to use instead 
a basis of eigenstates of $(L_3,K_3)$, which might give matrix elements and vertex
functions with the same classical limit. A detailed example of a small triangulation
is presented, which among other things  indicates, on the basis of a classical calculation,
that it would be  impractical to  limit oneself to tetrahedra with time-like normals.
}

\bigskip
\sect{ Introduction}
In the present formulation of spin-foam theory, as it has evolved from the 
original Barrett-Crane model \cite{BarrettCrane}, one associates  a discretized
 version of the Holst action \cite{Holst} $S_{Holst}=\int J^{IJ}\wedge F_{IJ}$ with a 
subdivision of space-time in 4-simplices; the curvature of the 
connection  $F_{IJ}$ is replaced by the product of the  SL(2,$\mathbb{C}$)
 holonomies along each `face' dual to a triangle (a `bone' in Regge's language).
For each triangle $t$ of a tetrahedron $\tau$ of the subdivision,  $J^{IJ}_t$  is 
 a linear combination of the area tensor  $S^{IJ}_t$ of the triangle
and of its Hodge dual ${ ^*\!S}^{IJ}_t=\half \epsilon^{IJKL}S_{tKL}$ 
that depends on a real parameter $\gamma$ or an angle $\theta$ \cite{DFLS}:
\be
J^{IJ}_t:={ ^*\!S}^{IJ}_t-\frac{1}{\gamma}S^{IJ}_t;\quad
\gamma=-i\,\frac{e^{i\theta}-1}{e^{i\theta}+1}\leftrightarrow
 e^{i\theta}=\frac{1+i\gamma}{1-i\gamma}
\label{holsttheta}\ee
The area tensors $S^{IJ}_t$ are constrained to be 'simple', the antisymmetrized 
product of two 4-vectors, and to satisfy 'closure', i.e. to sum to zero for 
each tetrahedron.

In the quantum theory the holonomies are in a representation of 
SL(2,$\mathbb{C}$) and the $J^{IJ}_t$ are identified with the generators of 
the  corresponding Lie Algebra.  In the spirit of Feynman's path integral,
integrating over the holonomies and summing/integrating over the representations 
should give a transition amplitude between an initial and a final configuration
of space. 
The discretization breaks e.g. invariance under diffeomorphisms, but
one may be optimistic, think of summing/refinining  triangulations, look
 at the successes of lattice QCD, etc.. 

A lot of work and a vast literature has been dedicated to the technical problems
of imposing the constraints in the quantum theory.  Part of the simplicity 
constraints are to be regarded as `second class', and the key idea of EPRL\cite{EPRL} 
is that  they should be imposed  on states \`a la Gupta-Bleuler'  \cite{LS}\cite{FK}.

There are however difficulties, and few detailed calculations
of a transition amplitude have so far been  attempted\cite{francesca}, while other discretized 
approaches, like Causal Dynamic Triangulation \cite{ambjorn} claim great success 
with large simulations. 
One difficulty that I see is that,
with the exception of \cite{Conrady3},  all the work carried on so far deals with
triangulations in which all tetrahedra are assumed to be space-like,
and an SU(2) basis is used for SL(2,$\mathbb{C}$) representations. 
This may be too restrictive; it makes sense to assume that all triangles
are space-like, without supposing that they all belong to 
space-like tetrahedra (i.e. with a time-like normal); this does not happen
classically in a simple example I give in \S 4.
One can follow \cite{Conrady3} \cite{Conrady1} and consider
time-like tetrahedra as well, but at the cost of using an SU(1,1) basis and facing
considerable tecnical difficulties. Besides, the idea that given a triangulation one 
should decide a priori which tetrahedra are time-like and which space-like is
unpleasant.  One would want an approach in which the distinction is simply 
 irrelevant \footnote{
In Causal Dynamic Triangulation \underline{all} triangles are assumed to be
timelike; a possible criticism is that the Courant criterium \cite{kitchensink} might then 
be systematically violated. This assumption is not modified 
in a recent paper\cite{renate} which relaxes the hypotheses on the foliation.}.  

In this note, I spell out the model in terms of spinor variables. 
Following \cite{DFLS},  a pair of spinors $(u_{t\alpha},t_{t\alpha})$ for each triangle
determine $J^{IJ}_t$. This gives a very neat way of expressing the constraints, 
with no distinction between time-like or space-like tetrahedra. However, in the details
 of the construction,  the difference does persist, although I sketch an approach
to some extent unified. In fact, it turns out that what makes a difference is 
whether a triangle is space- or time-like, i.e. whether its (area)$^2$  is 
positive or negative, and not to what type of tetrahedron  it belongs. 
Only positive  (area)$^2$ triangles can be described by 
spinor variables. In the quantum theory this implies that  the labels of the
representations of  SL(2,$\mathbb{C}$) contributing must be such that 
$\rho=\gamma n$, and that for triangles belonging to time-like tetrahedra one 
only need consider representations of SU(1,1) belonging to to the discrete series, with
labels  $k=\gamma\tfrac{n}{2}$.    Given the normal $V^I_\tau$ to the tetrahedron, 
the simplicity constraints imply for each triangle a relation between $u_t$ and 
$t_t$, very much like the `twistor equation' of Penrose. Using this relation one 
finds that  the light-like 4-vector $P^I_t=u_t\sigma^I\bar u_t$   
is orthogonal to  the area tensor  $S^{IJ}_t$ of the triangle, and that   
for each tetrahedron $V^I_\tau$ is proportional to the sum of the $P^I_t$ of
its four triangles, taken with signs $\kappa_t=\pm 1$. These signs must be such
that all $\kappa_tP_t^I$ are either incoming or outgoing, to satisfy the closure 
constraint. 

This representation of the geometry in terms of light-like 4-vectors coincides with
the picture conjectured by Yasha Neiman\cite{neiman}, that he correctly supposes
to be derivable from a spinor construction. And in the quantum theory
 I choose to realize the Hilbert spaces  ${\mathcal H}^{(n,\rho)}$
on which the representations of SL(2,$\mathbb{C}$) act as spaces of functions
of light-like 4-vectors (\cite{naimark}, pg.352). 

For triangles belonging to space-like tetrahedra, hence with time-like $V_\tau^I$,
one would choose the 'canonical basis' of ${\mathcal H}^{(n,\rho)}$, obtained 
reducing representations of SL(2,$\mathbb{C}$) with respect to its SU(2) subgroup, 
made  of eigenstates of $(\s L^2,L_3)$; or  for tetrahedra with space-like  $V_\tau^I$, 
the 'pseudobasis' obtained reducing with respect to SU(1,1),
 made of eigenstates of  the SU(1,1)  Casimir 
$Q=L_3^2-K_1^2-K_2^2,L_3$, but as I said considering only representations
of the discrete series. In both cases  
 the geometry, specifically the $P^I_t$, is encoded  using coherent states of the 
Perelomov type; these states are obtained applying appropriate rotations or boosts
to the 'highest weight' states; imposing the constraints 
 restricts the sums over the towers of representations to the lowest elements, 
$j=\tfrac{n}{2}$ or $k=\tfrac{n}{2}$.

In an attempt to treat the two cases in the same way I propose to
use a 'spinor basis', made of eigenstates of the generators $(L_3,K_3)$, with 
eigenvalues $(m,\lambda)$. 
The simplicity constraints imply that we use states 
such that  $m+\frac{1}{\gamma}\lambda=0$; applying to these states the same 
rotation or boost that we would have used to generate coherent states we obtain
states with the wanted $P^I_t$ as expectation value. So their direction is right, but
they do not have a definite area, as the eigenstates of $\s L^2$ or of $Q$; 
calculating the overlaps between these states and eigenstates of $(L_3,K_3)$
I find they  are peaked at  $\lambda\approx \tfrac{\rho}{2}$, but I cannot tell whether
this is enough to give the wanted behaviour in the semiclassical limit.  

This touches on a point which has given rise to some  debate:
 the asymptotic analysis of the `vertex function'\cite{barrettetal}, the building
 block of the transition amplitude, gives a behaviour $\sim \cos(area\cdot angle)$
 rather than the Regge $\sim e^{i\cdot area\cdot angle}$ one might have 
expected. Some time ago by D. Oriti and E. Livine \cite{oritilivine}pointed out 
that this means that in this form the model has no causality built in; although
causality is clearly what  distinguishes `euclidean' model, based on the group
O(4), from the proper lorentzian models, based on SL(2,$\mathbb{C}$). 
A detailed analysis traces this behaviour to the use of a first order
action\cite{riello}, and supports the idea that the implications of the $\cos(\ )$
term should be better understood, rather than corrected; other authors suggest ways to
modify the vertex function to get the  `correct' asymptotic behaviour  
\cite{engle}\cite{marko}.

To make the previous discussion more explicit I consider a specific  example in 
the last section of this note: the evolution of the simplest (5 points,
a `pentachoron') triangulation of $S^3$ from one time to the next. That results
 in a slice of spacetime made of 30  4-simplices, with faces  dual to 
either triangles with two vertices at time 0 and one at time 1, or viceversa. 
Looking closely at  one such face,  made of four steps, the question arises:
are all tetrahedra crossed space-like? classically, with the simplest uniform
assumptions, the answer turns out to be  no. I write down what would be the 
contribution of the face if one took  the spinor basis in all cases, and what
a reasonable justification for this choice could be. 

I have excluded the case of tetrahedra with a light-like normal, because including
it would have been a considerable complication, but see the very recent  
ref.\cite{simone}.  

\sect{spin-foam in terms of spinors}

 I use $\eta_{IJ}=(-+++),\ \epsilon^{0123}=1, \ a,b,...=1,2,3,\ 
\alpha,\beta,...=0,1$. More details about the notation used and various 
identities for spinors, 4-vectors and antisymmetric tensors have been collected into
 appendix A.

\subsection{ simplicity and closure.  }

I begin considering a 4-simplex with vertices $(abcde)$, bounded by tetrahedra
$(abcd), (abec)$, $(abde), (aced), (bcde)$, with flat metric $\eta_{IJ}$.
I shall label the triangles of the 
tetrahedron $(abcd)$: $abd, bac, cad, dbc$ as $t,s,..$. The triangle $t=(abd)$ 
is shared by $(abcd)$ and $(abde)$ and has an antisymmetric area tensor 
$S^{IJ}_t$, with Hodge dual   ${ ^*\!S}^{IJ}_t:=\half \epsilon^{IJKL}S_{t\,KL}$, 
(area)$^2$=$\half S^{IJ}_tS_{t\,IJ}$.

The components of any antisymmetric tensor, in particular of the 
generators of SL(2,$\mathbb{C}$) $J^{IJ}$, can be organized as two 3-vectors 
$L_a$ and  $K_a$, or their complex `left' and `right' combinations: 
\be
L_a:=\half\epsilon_{abc}J^{bc},\ K_a:=J^{0a};\quad J^{L,R}_a:=\half(L_a\pm iK_a),
\label{LK}\ee
or separating the `selfdual' and `antiselfdual' components:
\begin{align}
{ ^*\!J}^{IJ}:&=\half\epsilon^{IJKL}J_{KL}; \quad J^{IJ}_\pm:= \half(J^{IJ}\mp i\,
 { ^*\!J}^{IJ}):\ \half\epsilon^{IJKL}J_{\pm\, KL}=\pm i J^{IJ}_\pm;\nn\\ 
\half\epsilon_{abc}J^{bc}_\pm&=\pm iJ^{0a}_\pm= J_a^{L,R};
\quad J^{IJ}_\pm  J_{\pm IJ}=4J^{L,R}_aJ^{L,R}_a,\  J^{IJ}_\pm J_{\mp IJ}=0.
\label{Jpm}\end{align}
In terms of self-dual and antiself-dual components  $J^{IJ}_t$ becomes:
\be
J^{IJ}_t=
\frac{-1}{\sin\frac{\theta}{2}}(e^{i\frac{\theta}{2}}S^{IJ}_{t-}+
e^{-i\frac{\theta}{2}}S^{IJ}_{t+}),
\ee
and viceversa
\be
S^{IJ}_t=-\sin\frac{\theta}{2}(e^{-i\frac{\theta}{2}}J^{IJ}_{t-}+
e^{i\frac{\theta}{2}}J^{IJ}_{t+});  \quad
{ ^*\!S}^{IJ}_t=i\sin\frac{\theta}{2}(e^{-i\frac{\theta}{2}}J^{IJ}_{t-}-
e^{i\frac{\theta}{2}}J^{IJ}_{t+})
\label{SstarS}\ee
Tetrahedra being closed polyhedra, for each tetrahedron $\tau$ the area tensors 
must satisfy a closure constraint:
\be
\sum_{t\in\tau} S^{IJ}_t=0
\label{closure0}\ee
For each pair $t,s$ of triangles of the tetrahedron $(abcd)$ the quadratic 
'diagonal' and  `cross-diagonal'  `simplicity constraints' are expressed  as:
\begin{align}
{\rm 'diagonal'}\quad 
S^{IJ}_t{ ^*\!S}_{tIJ}&= i\sin^2\frac{\theta}{2}\big(e^{i\theta}J^L_{ta}J^L_{ta}-
   e^{-i\theta}J^R_{ta}J^R_{ta}\big)= 0,\quad\forall\; t\label{diag}\\
{\rm 'cross-diagonal'}\quad 
S^{IJ}_t{ ^*\!S}_{sIJ}&=i\sin^2\frac{\theta}{2}\big(e^{i\theta}J^L_{ta}J^L_{sa}-
e^{-i\theta}J^R_{ta}J^R_{sa}\big) =0,\quad   \forall \;t,s \label{crossdiag}
\end{align}
they are necessary and sufficient to guarantee that each $S^{IJ}_t$,
and therefore each  ${ ^*\!S}^{IJ}_t$, can be written as the
antisymmetrized product of two 4-vectors.
 A solution to these constraints is still a solution if we change 
$e^{i\theta}\to -e^{i\theta}$; this corresponds to switching 
$\gamma\to -\frac{1}{\gamma}$.

 A simpler, linear condition for simplicity is the existence, 
for each tetrahedron, of a 4-vector $V^I$, perpendicular to all its faces.
The proof is simple if $V_\tau^2\ne 0$: from the identity
\be
 { ^*\!S}^{IJ}_tV^M_\tau+{ ^*\!S}^{JM}_tV^I_\tau+{ ^*\!S}^{MI}_tV^J_\tau=
-\epsilon^{IJMN}S_{tNK}V^K_\tau
\label{theidentity}\ee
since  $V^2_\tau\ne 0$ and $V_{\tau M}S^{MI}_t=0$, it follows that:
\be
{ ^*\!S}^{IJ}_t=V^I\frac{V_M\,{ ^*\!S}^{MJ}_t}{V^2}-
\frac{V_M\,{ ^*\!S}^{MI}_t}{V^2}V^J :=V^IN^J_t-V^JN^I_t
\label{defN}\ee
and the dual relation $S^{IJ}_t=\epsilon^{IJKL}N_{tK}V_{\tau L}$, with $N^I_t$
ortogonal to  $V^I_\tau$ and to $S^{IJ}_t$.
So if $V^I$ exists,  ${ ^*\!S}^{IJ}_t$ is simple, and therefore so is $S^{IJ}_t$. 
$V^I$ necessarily exists for each tetrahedron,  because if for triangles 
$s, t:\ S^{IJ}_t=a^{[I}b^{J]}$,
$S^{IJ}_s=a^{[I}c^{J]}$, then  $V^I=\epsilon^{IJKL}a_Jb_Kc_L$ satisfies 
the assumptions.

From the definition follows that for any triangle
\be
{\rm (area)}^2=\half S^{IJ}_tS_{tIJ}=-\frac{1}{4} V^2_\tau N^2_t 
\nn\ee
This is negative for a time-like triangle, that can be brought to lie on the t-z plane, 
and then we can choose its normals to be either on the x or on the y axis;
positive for a space-like triangle, that can be brought to lie on the
 x-y plane, and then we can choose its normals to be either on the t or the z 
axis, and $V^2_\tau$ and $N^2_t$ will have opposite signs. I shall only deal with
this last case, for reasons that will be explained. 

One can also show that 
$S_{(abd)}^{IJ}=k\epsilon^{IJ}_{\ \ KL}V_{abcd}^K V_{abde}^L$, or a similar
expression in terms of the 4-vectors $N^I_t$,   $N'^I_t$ in the two neighbouring
tetrahedra.

The conclusion is that for each tetrahedron the constraints can be summarized as: 
\begin{align}
{\rm `closure':}&\ 
\sum_tN^I_t=\sum_t\frac{V_M\,{ ^*\!S}^{MI}_t}{V^JV_J}=
-\frac{i\sin\frac{\theta}{2} }{V^JV_J}
\sum_t\big(e^{i\frac{\theta}{2} }V_M J_{t+}^{MI}-
e^{-i\frac{\theta}{2}}V_MJ_{t-}^{MI}\big)=0 \label{closure}\\
{\rm `simplicity':}&\ 
V_MS^{MI}_t=-\sin\frac{\theta}{2}\,
\big(e^{i\frac{\theta}{2} }V_M J_{t+}^{MI}+
e^{-i\frac{\theta}{2}}V_MJ_{t-}^{MI}\big)= 0.\quad\forall\; t
\label{simplicity}
\end{align}
For space-like or time-like tetrahedra we can choose frames which give simple
explicit versions of (\ref{simplicity}):
\begin{align}
V^I=(V^0,0,0,0):\quad&L_1+\frac{1}{\gamma}K_1= L_2+\frac{1}{\gamma}K_2= 
L_3+\frac{1}{\gamma}K_3=0 \cr
V^I=(0,0,0,V^3):\quad&
L_1-\gamma K_1=L_2-\gamma K_2=L_3+\frac{1}{\gamma}K_3=0
\label{simplicity2}
\end{align}
 From  (\ref{closure}, \ref{simplicity}) combined  we get a simple form of the closure 
constraints for each tetrahedron, which do not depend on $V^I$:
\be
\sum_{t\in\tau} J_{ta}^R=0
\label{minclosure}\ee

\subsection{Using spinors}

Following  \cite{DFLS}, I shall use as basic   variables for spin-foam theory 
a SL(2,$\mathbb{C}$)  group element for each face, and a pair of spinors 
$(u_t,t_t)$ for each triangle. For SL(2,$\mathbb{C}$) transformations $u$ and $t$
  transform as\footnote{ since I have SL(2,$\mathbb{C}$) transformations 
acting on the right, many of the definitions that follow will be different in
detail from \cite{DFLS}.} :
\be
u\to u g \quad t \to  tg^{\dagger-1}
\label{trfspinors}\ee
The idea is to implement these transformations through the  SL(2,$\mathbb C$)
invariant Poisson brackets $\{t_\alpha,\bar u_\beta\}=-i\delta_{\alpha\beta};\quad
\{u_\beta,\bar t_\alpha\}=-i\delta_{\alpha\beta}$ and  the generators:
\be
J^L_a=-\half u\sigma_a\bar t,\quad J^R_a=-\half t\sigma_a\bar u;\quad
\{J^{L,R}_a,J^{L,R}_b\}=\epsilon_{abc}J^{L,R}_c,\quad\{J^L_a,J^R_b\}=0.
\label{defJLJR}\ee
so that 
\be
\{u_\alpha,J^L_a\}=\ihalf u_\beta\sigma_{a\beta\alpha};\quad
\{t_\alpha,J^R_a\}=\ihalf t_\beta\sigma_{a\beta\alpha}
\nn\ee
 Altogether, with  a pair of  spinors $(t_t,u_t)$  for each of the 4  
triangles of a tetrahedron we define  $\s L_t,\ \s K_t,\ \s J_t^{L,R}$,
to which we want to impose the closure and simplicity constraints.
 
The closure constraints give for each tetrahedron $\tau$,  from (\ref{minclosure}):
\be 
\sum _{t\in\tau}\bar u_{t\alpha}\sigma_{a\alpha\beta}t_{t\beta}=0
\label{minclosureb}\ee
The matrix $\sum_{t\in\tau} \bar u_{t\alpha}t_{t\beta}$ must therefore be
 proportional to the unit matrix, or zero, hence:
\be
\sum_{t\in\tau}\bar u_{t\alpha}t_{t\beta}=\half \sum_{t\in\tau}(u_t^\dagger t_t)
\delta_{\alpha\beta}:=C_\tau e^{i\psi_\tau}\delta_{\alpha\beta},
\label{clo}
\ee
with $C_\tau$ a real constant. We shall see shortly that for all tetrahedra
$e^{i\psi_\tau}=e^{i\frac{\theta}{2}}$.

The relations  (\ref{simplicity}) expressing the simplicity constraints, multiplied
  by $\tilde\sigma_I=(1,\sigma_a)$ and  dropping overall constants become,
using (\ref{spinoridb}, \ref{defJLJR}):
\be
e^{i\frac{\theta}{2}}\tilde\sigma_IV^I\sigma_a  (u_t\sigma_a \bar t_t)-
e^{-i\frac{\theta}{2}}\sigma_a (t_t\sigma_a\bar u_t )\,\tilde\sigma_IV^I=0,
\label{simplicitybis}\ee
where 
\[\sigma_a(u\sigma_a\bar t)=
\left(\begin{matrix}
 u_0\bar t_0- u_1\bar t_1 & 2 u_1\bar t_0\\2u_0\bar t_1& u_1\bar t_1- u_0\bar t_0
       \end{matrix}\right)
\]
For either spacelike or timelike tetrahedra we can go to a standard frame 
in which
\be
V^I_0=(\alpha,0,0,\beta)
\label{vzero}\ee
excluding the light-like case $\alpha^2=\beta^2$. Indicating the 
spinors in this frame with capital letters, from  (\ref{simplicitybis}) we 
get the independent equations: 
\be
(\alpha+\beta)e^{i\frac{\theta}{2}}U_1\bar T_0=e^{-i\frac{\theta}{2}}(\alpha-\beta)T_1\bar U_0;
\quad 
e^{i\frac{\theta}{2}}(U_0\bar T_0-U_1\bar T_1)=e^{-i\frac{\theta}{2}}T_0\bar U_0-T_1\bar U_1.
\label{UTeqs}\ee
in fact, three independent real equations. One can use them
to get an explicit relation between the $U$ and the $T$ spinors for each
triangle. In fact these equations are  satisfied if we set:
\be
T_0=\kappa(\beta-\alpha)e^{i\frac{\theta}{2}}U_0,\ 
T_1=\kappa(\beta+\alpha)e^{i\frac{\theta}{2}}U_1\ \to\
T_\alpha=\kappa e^{i\frac{\theta}{2}}U_\beta\sigma_{I \beta\alpha}V^I_0 
\label{thesolution}\ee
 where $\kappa$ could be any real number, but for simpliciy I choose
$\kappa=\pm 1$.
We can  go back to an arbitrary frame by boosting  
$(V^I_0, T_t,U_t)$ to $(V^I,t_t,u_t)$ with some $g\in SL(2,\mathbb{C})$. 
From (\ref{trfvectors}), (\ref{trfspinors}) we have:
\be 
T_\alpha=t_\beta\, g^{\dagger -1}_{\beta\alpha};\quad U_\alpha=u_\beta\, g_{\beta\alpha}
;\quad V_0^I=V^J\Lambda_J^{\ I}
\label{boost}\ee
and the relation between these 'boosted' spinors will be:
 \be
t_{t\alpha}-\kappa_te^{i\frac{\theta}{2}}\,u_{t\beta}\,\sigma_{I\beta\alpha}V^I=0.
\label{solution}\ee
They will 'solve' the constraint eq.s (\ref{simplicitybis}), because these equations are 
covariant.
Eq. (\ref{solution}) is the main result of this section; it is interesting 
to notice that apart from factors it is the `twistor equation' of  R. Penrose.

The first consequence we want to draw from (\ref{solution})  comes from looking at
 the normals to the triangles (\ref{defN}), for which, if the (\ref{simplicitybis}) are 
satisfied, one finds:
\be
\tilde\sigma_IN^I_t=\frac{\tilde\sigma_IV_{\tau M}\,{ ^*\!S}^{MI}_t}{V_\tau^2}=
-\frac{\sin\frac{\theta}{2}}{2V^2_\tau}
\big(e^{i\frac{\theta}{2}} \tilde\sigma_IV^I_\tau\sigma_a (u_t \sigma_a\bar t_t)
       +e^{-i\frac{\theta}{2}} (t_t\sigma_a\bar u_t)\sigma_a \tilde\sigma_IV^I_\tau\big)
\label{normal}\ee
Using (\ref{solution}) this gives, after some algebra, the remarkably simple expressions:
\be
N^I_t = \frac{\kappa_t\sin\frac{\theta}{2}} {V^2_\tau}
(V^I_\tau V^J_\tau-\eta^{IJ}V^2_\tau)u_t\sigma_J\bar u_t;\quad
N^2_t=- \frac{\sin^2\frac{\theta}{2}} {V^2_\tau}(u_t\sigma_IV^I_\tau\bar u_t)^2
\label{normal2}\ee
  In two particular cases we have:  

for $V^I_\tau=(V^0,0,0,0)$, 
$N^I_t=\kappa_t\sin\frac{\theta}{2}\big(0,-(u_t\sigma_a\bar u_t)\big)$;

for $V^I_\tau=(0,\s v)$, $N^I_t=\kappa_t\sin\frac{\theta}{2}
\big(-u_t\bar u_t, \frac{1}{\s v^2}(v_av_b-\s v^2\delta_{ab})
(u_t\sigma_b\bar u_t)\big)$.\\
In general, from the definition of $N^I_t$ we get: 
\be
S^{IJ}_t=\epsilon^{IJKL}N_{tK}V_{\tau L}=k_t\sin\frac{\theta}{2}\,\epsilon^{IJKL}V_{\tau K}
\,u_t\sigma_L\bar u_t
\ee
it follows that the 4-vector $P_t^I:=u_t\sigma^I \bar u_t$,
 which  is null, is orthogonal to $S^{IJ}_t$:
\be
S^{IJ}_t u_t\sigma_I \bar u_t=S^{IJ}_tP_{tI}=0.
\label{orthoP}\ee
and must therefore be in the plane spanned by 
$V_\tau^I$ and $N^I_t$; but this implies that $V_\tau^2$ and $N^2_t$ have
opposite signs, and indeed from (\ref{normal2}) we see that $V^2_\tau N_t^2$  
 is \underline{ always negative},
 and therefore ${\rm (area)}^2=\half S^{IJ}_tS_{tIJ}>0$. 
I would conclude that a configuration in which this does not happen cannot be
 described by spinors the way we have  introduced them. 

From the expression of $N^I_t$ for $V^I_\tau=(0,\s v)$ we see that for time-like
 tetrahedra the $\kappa_t$ cannot be all of the same sign, because this would clash with 
the closure constraint $\sum_{t\in\tau}N^0_t=0$. 

Combining (\ref{clo}) with (\ref{solution}) we find that if $C_\tau\ne 0$:
\be
\big(\sum_{t\in \tau}\bar u_{t\alpha}\kappa_t e^{i\frac{\theta}{2}}u_\gamma\big)
\sigma_{I\gamma\beta}V^I_\tau=C_\tau e^{i\frac{\theta}{2}}\delta_{\alpha\beta}
\label{inverse1}\ee
In this way we find the inverse of $\sigma_IV^I_\tau$; provided that 
$V^2_\tau\ne 0$ and $C_\tau\ne 0$ we can write the 4-vector normal to the 
tetrahedron as 
\be
V^I_\tau=\frac{V^2_\tau}{2C_\tau}\sum_{t\in\tau}\kappa_t\, u_t \sigma^I\bar u_t=
\frac{V^2_\tau}{2C_\tau}\sum_{t\in\tau}\kappa_t P^I_t
\label{inverse2}\ee

Finally from  (\ref{solution}), for any pair of 
triangles $t,s$ belonging to  a tetrahedron with normal $V^I_\tau$ and spinors
$(u_s,t_s), (u_t,t_t)$, we can derive, using
$\sigma_IV^I\tilde\sigma_JV^J=V^2$ and the identities in (\ref{trfvectors}):
\begin{align}
t_{t\alpha}\epsilon_{\alpha\beta}t_{s\beta}&+
\kappa_s\kappa_tV^2_\tau e^{i\theta}
\,u_{t\alpha}\epsilon_{\alpha\beta}u_{s\beta}=0;
\label{holo}\\
t_{t\alpha }\bar u_{t\alpha}&=
\kappa_t e^{i\frac{\theta}{2}}u_{t\alpha}\,\sigma_{I\alpha\beta}V^I\bar u_{t\beta}=
e^{i\theta}\bar t_{t\alpha } u_{t\alpha}
\label{holobis}\end{align}
eq.(\ref{holobis}) implies $e^{i\psi_\tau}=e^{i\frac{\theta}{2}}$, as anticipated.

Eq.s  (\ref{holo}) are the `holomorphic simplicity 
constraints' of  \cite{DFLS}; they form a set of first class constraints, but  
one can see that they imply (\ref{diag}) and the second class 
(\ref{crossdiag})\footnote{notice that (\ref{solution}) and (\ref{holo}) 
\underline{do not} Poisson-commute with their complex conjugates}:
in fact, from the definition (\ref{defJLJR}),  using (\ref{spinorida}) we have:
\be
\s J_t^L\cdot\s J_s^L =\quart t_{t}^\dagger u_{t}\;t_{s}^\dagger u_{s}-
\half \bar t_{t\alpha}\epsilon_{\alpha\gamma} \bar t_{s\gamma}\,
u_{t\beta}\epsilon_{\beta\delta} u_{s\delta},\quad
\s J_t^R\cdot\s J_s^R=\quart  u_{t}^\dagger t_{t}\;u_{s}^\dagger t_{s}-
\half  \bar u_{t\alpha}\epsilon_{\alpha\gamma}  \bar u_{s\gamma}\,
t_{t\beta} \epsilon_{\beta\delta}t_{s\delta}.\label{jrjl}
\ee
and from (\ref{holo}):
\be
 e^{i\theta}t_{t}^\dagger u_{t}\;t_{s}^\dagger u_{s}=
 e^{-i\theta}u_{t}^\dagger t_{t}\;u_{s}^\dagger t_{s};\quad
e^{i\theta}\bar t_{t\alpha}\epsilon_{\alpha\beta}\bar t_{s\beta}\,
 u_{t\gamma}\epsilon_{\gamma\delta}u_{s\delta}=
e^{-i\theta}t_{t\alpha}\epsilon_{\alpha\beta}t_{s\beta}\,
\bar u_{t\gamma}\epsilon_{\gamma\delta}\bar u_{s\delta}, \label{sixtyfour}
\ee
from which (\ref{diag})(\ref{crossdiag}) follow.

 Overall, in this scheme the triangle $t=(abd)$  
is shared by $(abcd)$ and $(abde)$ within a 4-simplex $(abcde)$; these tetrahedra
have different normal vectors $V^I_{abcd}, V^I_{abde}$ and the triangle
 different normals, but the same $S^{IJ}_t$ and spinors $(u_t,t_t)$.

\sect{The quantum theory}

\subsection{Representations of SL(2,$\mathbb{C}$)}

 In the quantum theory the constraints should be imposed by restricting the 
sum/integral over the representations of SL(2,$\mathbb{C}$), an idea that goes back to the 
orignal  formulation of spin-foam theory\cite{BarrettCrane}, 
although the breakthrough
 in the theory came much later, with the EPRL \cite{EPRL} idea that this restriction should 
be derived imposing the constraints on the states, \`a la Gupta-Bleurer 
\footnote{ \cite{Conrady3} has a particularly lucid explanation.}.

Given a triangulation, the contribution of a face to the transition amplitude is 
determined
by a sequence of transitions between `states'. These states  are elements of 
${\mathcal H}^{(n,\rho)}$, the Hlbert space on which a unitary representation of 
SL(2,$\mathbb{C}$) acts; they encode the geometry of the tetrahedron
involved, and must be such that the expectation of $J^{IJ}$ satisfies 
the simplicity constraints. 

 The theory of unitary representations of 
SL(2,$\mathbb{C}$) \cite{naimark} is  the work of I. M. Gel'fand and his collaborators; 
the representations are infinite dimensional, labeled by  the two indices 
$(n,\rho)$ with $n$ integer, $\rho$ real $\in(-\infty,\infty)$. On functions of 
a spinor $u=(u_0\ u_1)$ and its c.c., homogeneous of degree 
$(\lambda,\mu)=(\frac{n}{2}+i\frac{\rho}{2}-1,-\frac{n}{2}+i\frac{\rho}{2}-1)$,
they act as $T_g^{(n\rho)}f(u,\bar u)=f(u\cdot g,\bar u\cdot\bar g)$.
With the measure   
$\Omega_u=\frac{i}{2}(u_0 du_1-u_1du_0)\wedge(\bar u_0d\bar u_1-\bar u_1d\bar u_0)$ 
these functions form the Hilbert space ${\mathcal H}^{(n,\rho)}$.
There are however other realizations of ${\mathcal H}^{(n,\rho)}$, as we shall see
in a moment.

A simple version of Gelfand's construction has been given by Smorodinskii and Huszar
\cite{smorodinski}\cite{huszar}, based on the idea of separating right and left 
generators of  SL(2,$\mathbb{C}$), extending the theory of representations of
 SU(2) with  $J^L_3,J^R_3$ diagonal; the unitarity of the representation 
requires $(J_a^L)^\dagger=J_a^R$.
By simple algebraic manipulations\footnote{and demanding that the group 
representation be one valued} they find that 
 the Casimir operators for the $(n,\rho)$ representation can be expressed as:
\be
(\s J^L)^2=\quart(\frac{n}{2} -i\frac{\rho}{2})^2-\quart ; 
\quad (\s J^R)^2=\quart(\frac{n}{2} +i\frac{\rho}{2})^2-\quart  
\label{casimir}\ee
equivalent to the more common:
\be
C_1=J^{IJ}J_{IJ}=2(\s L^2-\s K^2)=\half(n^2-\rho^2-4),\quad
C_2=J^{IJ}{ ^*\!J}_{IJ}=-4\s L\cdot\s K=n\rho
\label{casimir1}\ee
These expressions immediately suggest that in a quantum theory, for large $n$ and  
ignoring the $\quart$, the constraint (\ref{diag}) will be satified if we limit
the sum over representations of SL(2,$\mathbb{C}$) to those such that:
\be
e^{i\theta}(\s J^L)^2= e^{-i\theta}(\s J^R)^2\ \to\
\frac{(n+i\rho)^2}{(n-i\rho)^2}=e^{2i\theta}
=\frac{(1+i\gamma)^2}{(1-i\gamma)^2}
\label{casimir2}\ee
As already pointed out after eq.s (\ref{diag},\ref{crossdiag}), this equation has two
 solutions: $\rho=n\gamma$, or $\rho=-\frac{n}{\gamma}$,
that correspond to the two choices $\pm e^{i\theta}$ for the square root of 
$e^{2i\theta}$. But now we can see that positive (area)$^2$, the only case we want to deal with,
correspond to the choice $\rho=\gamma n$; In fact, with some algebra, I find that
(\ref{SstarS}) and (\ref{casimir2}) imply that:
 \begin{align}
{\rm (area)}^2&=\half S^{IJ}S_{IJ}=\frac{\gamma^2}{2(1+\gamma^2)^2}
((1-\gamma^2)C_1+2\gamma C_2)\cr
\rho=\gamma n\to {\rm (area)}^2=\tfrac{1}{4}\gamma^2n^2&-
\frac{\gamma^2(1-\gamma^2)}{2(1+\gamma^2)^2};\quad
\quad\rho=-\frac{n}{\gamma} \to {\rm (area)}^2=-\tfrac{1}{4}n^2
-\frac{\gamma^2(1-\gamma^2)}{(1+\gamma^2)^2}
\label{areaquant}\end{align}
Therefore choosing $\rho=\gamma n$ one has that, up to terms which do not grow with
 $n$, {\rm (area)}$^2$ is positive and quantized, one of the key results of loop  
quantum gravity.

With the same algebra\footnote{given as problem 2.9 in \cite{carmeli}!} 
that leads to (\ref{casimir}) one defines a `canonical basis' for the Hilbert space 
${\mathcal H}^{(n,\rho)}$;  it consists of eigenstates  $\psi^j_m$ of 
$(\s L^2, L_3)$ 'injected' in ${\mathcal H}^{(n,\rho)}$, such that in particular:
\begin{align}
\s L^2\psi^{(n,\rho) j}_m=&j(j+1)\psi^{(n,\rho) j}_m;\ L_3 \psi^{(n,\rho) j}_m=m\psi^{(n,\rho) j}_m ;\ 
\ L_\pm \psi^{(n,\rho) j}_m=\sqrt{(j\pm m+1)(j\mp m)} \psi^{(n,\rho) j}_{m\pm 1};\cr
K_3|\psi^{(n,\rho) j}_m>&=C_j|\psi^{(n,\rho) j-1}_m>-
\frac{\rho n m}{4j(j+1)}|\psi^{(n,\rho) j}_m>-C_{j+1}|\psi^{(n,\rho) j+1}_m>.
\label{k3actsu2}\end{align}
(the coefficients $ C_j$ depend on $\rho,n,m$ besides $j$; we shall not need their
form).

I shall use a realization of ${\mathcal H}^{(n,\rho)}$   as a space of 
functions on light-like 4-vectors $P^I$ \cite{lomont}\cite{huszar}
\cite{naimark},
derived from a spinor $u_\alpha$ by $P^I:=u\sigma^I\bar u$. 
This choice is in principle convenient because
for  any triangle $t$ we have a null 4-vector $P_t^I=u_t\sigma^I\bar u_t$, and we
have seen earlier (\ref{orthoP}) that the simplicity constraints imply that $P_t^I$
 is orthogonal to $S^{IJ}_t$. To get an element of ${\mathcal H}^{(n,\rho)}$ we 
start from a function $\Phi(P)$ such that 
$\Phi(e^\alpha P)=e^{(i\frac{\rho}{2}-1)\alpha}\Phi(P)$, and set
\be
f(u_0,u_1)= 
(\frac{u_0}{\bar u_0})^{\frac{n}{2}} \Phi(P)
\label{deff}\ee
From this definition follows that under SL(2,$\mathbb{C}$) transformations,
 from(\ref{trfvectors}):
\be
T^{(n\rho)}_g\Big((\frac{u_0}{\bar u_0})^{\frac{n}{2}} 
\Phi( u\sigma^I\bar u)\Big)=
\big(\frac{(ug)_0}{(\bar u\bar g)_0}\big)^{\frac{n}{2}} 
\Phi(u g\sigma^I g^\dagger\bar u)=
\big( \frac{(ug)_0}{(\bar u\bar g)_0}\big)^{\frac{n}{2}} \Phi(P^J\Lambda_J^{\ I})
\label{trfP}
\ee
The natural invariant scalar product would be 
\be
<\Psi|\Phi>=\int \theta(P^0)\delta(P^2)\frac{d^4P}{(2\pi)^3}
\,\overline{\Psi(P)}\, \Phi(P)=\int\frac{d^3P}{2(2\pi)^3P}\,\overline{\Psi(P)}\,\Phi(P)
\label{scapro}\ee
For infinitesimal transformations 
$g(\epsilon,\eta)=1+\frac{1}{2}(i\epsilon_a+\eta_a)\sigma_a
\simeq (1+i\epsilon_aL_a+i\eta_aK_a)$:
\begin{align}
T^{(n\rho)}_{g(\epsilon,\eta)}\Big((\frac{u_0}{\bar u_0})^{\frac{n}{2}} 
\Phi( u\sigma^I\bar u)\Big)&=
\Big(\frac{u_0+\tfrac{1}{2}u_0(i\epsilon_3+\eta_3)+
\tfrac{1}{2}u_1(i\epsilon_1+\eta_1-\epsilon_2+i\eta_2) }
{\bar u_0+\tfrac{1}{2}\bar u_0(-i\epsilon_3+\eta_3)+
\tfrac{1}{2}\bar u_1(-i\epsilon_1+\eta_1-\epsilon_2-i\eta_2)}\Big)^{\frac{n}{2}}
\cdot\cr
 &\quad\quad\cdot\Phi\big( u_\alpha\sigma^I_{\alpha\beta} \bar u_\beta
+\tfrac{i}{2}\epsilon_a u [\sigma_a,\sigma^I] \bar u
+\tfrac{1}{2}\eta_a u\{\sigma_a,\sigma^I\}u\big)=\cr
&=(\frac{u_0}{\bar u_0})^{\frac{n}{2}}(1+i\sum(\epsilon_kL_k+\eta_kK_k))
\Phi(P);
\end{align}
expanding, we have for the generators of SL(2,$\mathbb{C}$)
\cite{lomont}\cite{huszar}:
\begin{align}
L_1&=-i(P_2\partial_3-P_3\partial_2)+\frac{n P_1}{2(P+P_3)},\quad
L_2=-i(P_3\partial_1-P_1\partial_3)+\frac{n P_2}{2(P+P_3)}, \cr
L_3&=-i(P_1\partial_2-P_2\partial_1)+\frac{n }{2}\label{thegenerators}\\
K_1&=-iP\partial_1-\frac{n P_2}{2(P+P_3)},\quad 
K_2=-iP\partial_2+\frac{n P_1}{2(P+P_3)},\quad K_3= -iP\partial_3,
\nn\end{align}
where  $P^0=P=\sqrt{P_1^2+P_2^2+P_3^2}$, and
$\partial_a\Phi:=\frac{\partial\Phi}{\partial P_a}\big|_{P^0}
+\frac{\partial\Phi}{\partial P^0}\big|_{P_a}\frac{\partial P}{\partial P_a}$.

If one takes $P^I=e^\alpha(a,b\cos\varphi,b\sin\varphi,c),\ a=\sqrt{b^2+c^2}$, then:
\be
\frac{\partial}{\partial \alpha}=P_a\partial_a;\qquad
\frac{\partial}{\partial\varphi}= P_1\partial_2-P_2\partial_1\quad \rightarrow\quad
L_3=-i\frac{\partial}{\partial \varphi}+\frac{n}{2}
\label{genericbasis}\ee
For $\Phi\in{\mathcal H}^{(n,\rho)}, \frac{\partial\Phi}{\partial \alpha}=
(i\frac{\rho}{2}-1)\Phi$;  therefore there is no information in the dependance of $\Phi$ 
on $\alpha$. Apart from a factor $e^{(i \frac{\rho}{2}-1)\alpha}$ the homogeneous 
function  $\Phi$ is completely determined by its value on the intersection of the
light-cone with the sphere $P=e^{\alpha}$, or with the plane $P^3=e^\alpha$,
or with the hyperbola $(P^0)^2-(P^3)^2= e^{2\alpha}$.  I shall consider three basis
adapted to each choice;  in each case for the scalar product (\ref {scapro})
should be multiplied by an appropriate $\delta$ function, or the integration in
$d\alpha$ over the generators of the light-cone simply omitted.                                                                                                                                                                                                                                                                          


\subsection{The various basis of ${\mathcal H}^{(n,\rho)}$.}

To realize explicitely the canonical basis I specialize further the choice of
 variables choosing:
\be
u=e^{\frac{a}{2}}(\cos\frac{\theta}{2}e^{i\frac{\varphi}{2}},
\sin\frac{\theta}{2}e^{-i\frac{\varphi}{2}}), \quad
P^I=e^a(1,\sin\theta\cos\varphi,\sin\theta\sin\varphi,\cos\theta).
\label{L2}\ee 
To impose the simplicity constraints we use our knowledge of action of $K_3$ on 
the basis states  of the canonical basis  (\ref{k3actsu2}).
 Choosing $V^I=(V^0,0,0,0)$, we find  \cite{Conrady3} for the component 3 of 
 the constraint (\ref{simplicity2}):  
\be
<\psi^{(n\rho)j}_{m'}|  (L_3+\frac{1}{\gamma}K_3)|\psi^{(n\rho)j}_{m}>=
\delta_{mm'}(m-\frac{\rho nm}{4j(j+1)\gamma})
\label{expectationk}\ee
If $\rho=\gamma n$, this will be approximately 0 if we take $j= \frac{n}{2}$. Then
the expectation values of components 1 and 2 will also vanish, because
\be
L_\pm+\frac{1}{\gamma}K_\pm=\pm [L_3+\frac{1}{\gamma}K_3,L_\pm]
\label{commrel1}\ee
where $L_\pm=L_1\pm iL_2, \ K_\pm=K_1\pm iK_2$.
This $\rho=n\gamma,\ j= \frac{n}{2}$ solution on the other hand matches very
nicely with the quantization of areas in loop quantum gravity. In fact, 
since $\rho=\gamma n$, the constraint implies:
\begin{align}
(\s L+\frac{1}{\gamma}\s K)^2&=(1+\gamma^2)\s L^2-\half C_1-\half\gamma C_2=
(1+\gamma^2)(\s L^2-\quart n^2)+1\simeq 0\cr
&\to\ {\rm (area_t)}^2=\gamma^2\s L^2=\gamma^2j(j+1)
\end{align}
To find the corresponding basis functions I use the expresson of $\s L^2$ in
 these coordinates and a property of the Wigner functions:  
\begin{align}
\s L^2&=-\frac{1}{\sin\theta}\frac{\partial}{\partial\theta}\sin\theta
\frac{\partial}{\partial\theta}-\frac{1}{\sin^2\theta}\frac{\partial^2}{\partial\theta^2}
+\frac{n}{1+\cos\theta},\cr
\Big(-\frac{1}{\sin\theta}\frac{\partial}{\partial\theta}
\sin\theta\frac{\partial}{\partial\theta}&+\frac{1}{\sin^2\theta}
(m^2+m'^2-2mm'\cos\theta)\Big)\,d^j_{mm'}(\theta)=j(j+1)d^j_{mm'}(\theta).
\label{propwig}\end{align}
One finds \cite{huszar} the basis functions that diagonalize $(\s L^2, L_3)$:
 \begin{align}
|\psi^{(n\rho)\frac{n}{2}}_{m}>\ &\leftrightarrow\ 
\psi_{m}^{(n\rho)\frac{n}{2}}(a,\theta,\varphi)=C_ae^{(i\frac{\rho}{2} -1)a}
e^{i(m-\frac{n}{2})\varphi}d^\frac{n}{2}_{m\frac{n}{2} }(\theta),\cr
C_a=2\pi\sqrt{n+1};\quad&  -\tfrac{n}{2}\le m\le\tfrac{n}{2} ;\quad
<\psi_{m'}^{(n\rho)\frac{n}{2} }|\psi_{m}^{(n\rho)\frac{n}{2} }>=\delta_{mm'}
\label{canbasis}\end{align}

Coherent states carry the information on the geometry of a given triangle,
 in this case on the normal $\s n=\frac{\s N}{|\s N|}=(\sin\bar\theta\cos\bar\varphi,
\sin\bar\theta\sin\bar\varphi,\cos\bar\theta)$; they
are defined applying to the `highest weight' state 
$|\psi^{(n\rho)\frac{n}{2}}_{\frac{n}{2}}>$ the rotation:
\begin{align}
u_{\s n}&=e^{-i\frac{\sigma_3}{2}\bar\varphi}e^{-i\frac{\sigma_2}{2}\bar\theta};\quad
u_{\s n}^\dagger\left(\begin{matrix}\sigma_1\\\sigma_2\\\sigma_3\end{matrix}\right)u_{\s n}=
\left(\begin{matrix}
\cos\bar\theta\cos\bar\varphi&-\sin\bar\varphi&\sin\bar\theta\cos\bar\varphi\\
\cos\bar\theta\sin\bar\varphi&\cos\bar\varphi&\sin\bar\theta\sin\bar\varphi\\
-\sin\bar\theta&0&\cos\bar\theta\end{matrix}\right)\left(\begin{matrix}\sigma_1\\
\sigma_2\\\sigma_3\end{matrix}\right).\cr
&|\psi^{(n\rho)\frac{n}{2}}_{ \s n}>:={\cal D}^{\frac{n}{2}}(u_{\s n})
 |\psi^{(n\rho)\frac{n}{2}}_{\frac{n}{2}}>;\quad
 <\psi^{(n\rho)\frac{n}{2}}_{\s n}|L_a |\psi^{(n\rho)\frac{n}{2}}_{\s n}>=\tfrac{n}{2}\,n_a
\label{cohesu2}\end{align}
These states are of minimal uncertainty; for example, indicating expectation
 values on them simply as $<.>$ we have
\be
\frac{\Delta L}{<|L|>}=\frac{\sqrt{<(L_a-<L_a>)(L_a-<L_a>)>}}{\sqrt{<L_aL_a>}}=
\frac{\sqrt{j(j+1)-j^2}}{\sqrt{j(j+1)}}\simeq\frac{\sqrt{j}}{j}
\nn\ee

For time-like tetrahedra one would choose $V^I=(0,0,0,\pm 1)$; its little group is
SU(1,1), the set of all $v\in\,$SL(2,$\mathbb{C}$) such that 
$v^\dagger\sigma_3v=\sigma_3$. With its unitary irreducible representations
\cite{bargmann}
one defines a `pseudobasis' \cite{Ruhl} for  ${\mathcal H}^{(n,\rho)}$.  The reduction of 
${\mathcal H}^{(n,\rho)}$ to representations of SU(1,1) has been studied very 
thoroughly by F. Conrady and  J. Hnybida \cite{Conrady3},\cite{Conrady1},\cite{Conrady2},,
 and before them by many authors, beginning with  A. Sciarrino and M. Toller
\cite{sciatol}\cite{Ruhl}\cite{mukunda}.  SU(1,1)  has infinite dimensional 
unitary representations labeled by the  eigenvalue of its Casimir operator 
$Q=L_3^2-K_1^2-K^2_2$, which has a discrete spectrum $q=k(k-1), \ k$ (half)integer,
 hence positive q, and a continous one $q=k(k+1),\ k=-\half+is,\ 0<s<\infty$, 
hence negative q; eigenstates of $(Q, L_3)$  injected in ${\mathcal H}^{(n,\rho)}$
 form altogether the pseudobasis. Choosing variables: 
\begin{align}
u&=(e^{\frac{b}{2}} \cosh\frac{t}{2}e^{i\frac{\varphi}{2}},
e^{\frac{b}{2}} \sinh\frac{t}{2}e^{-i\frac{\varphi}{2}}),\quad
P^I=(e^b\cosh t,e^b\sinh t\cos\varphi,e^b\sinh t\sin\varphi, e^b);\cr
Q&=L_3^2-K_1^2-K^2_2=
\frac{1}{\sinh t}\frac{\partial}{\partial t}\sinh t\frac{\partial}{\partial t}+\
\frac{1}{\sinh^2 t}\frac{\partial^2}{\partial\varphi^2}+n\frac{1}{1+\cosh t}L_3.
\nn\end{align}
We are  only concerned with the states in the discrete spectrum,
because spinors are  suitable only for positive (area)$^2$, hence $Q>0$ 
representation. In fact, using the constraints and $\rho=\gamma n$ as 
above \cite{Conrady3}, we find for the quantized areas:
\begin{align}
(L_1-\gamma K_1)^2+&(L_2-\gamma K_2)^2-(K_3+\gamma L_3)^2=
\half C_1+\half\gamma C_2-(1+\gamma^2)Q=\cr
&=(1+\gamma^2)(\quart n^2-Q)-1
\simeq 0 \to {\rm (area)}^2\simeq\gamma^2 Q= \gamma^2k(k-1)
\end{align}
  For the discrete series, the analogues of (\ref{canbasis}, \ref{k3actsu2}) 
are\cite{mukunda}:
\begin{align}
Q|\psi^{(n,\rho) k\pm}_m>&=k(j-1)|\psi^{(n,\rho) k\pm}_m>;
\quad L_3 |\psi^{(n,\rho) k\pm}_m>=m|\psi^{(n,\rho) k\pm}_m >;\cr 
(K_2\mp iK_1)|\psi^{(n\rho)k\pm}_m>&=\sqrt{(m\mp k+1)(m\pm k)}|\psi^{(n\rho)k\pm}_{m\pm 1}>\cr
K_3f^{(n\rho)k\pm}_m=A^{(n\rho)k}_m&f^{(n\rho)(k+1)\pm}_m -
\frac{\rho mn}{4k(k-1)}|\psi^{(n\rho)k\pm}_m >+
A^{(b\rho)k-1}_m|\psi^{(n\rho)(k-1)\pm}_m>
\label{k3actsu11}\end{align}    
(we do not need the expression of $A^{(n\rho)k}_m$).
If $V^I=(0,0,0,1)$, the expectation value of  
component 3 of  the constraint (\ref{simplicity2}) is given by: 
\be
<\psi^{(n\rho)k\pm}_{m'}|(L_3+\frac{1}{\gamma}K_3)|\psi^{(n\rho)k\pm}_m>=
\delta_{mm'}(m-\frac{\rho nm}{4k(k-1)\gamma})
\nn\ee
which, given that $\rho=\gamma n$, will vanish if $k=\frac{n}{2}$. Then
the expectation values of components 1 and 2 will also vanish because:
\be
K_\pm-\frac{1}{\gamma}L_\pm= \pm([L_3+\frac{1}{\gamma}K_3,K_\pm])
\label{commrel2}\ee
In conclusion, the states $|\psi^{(n\rho)k\pm}_m>$
 of the discrete spectrum that will contribute are those with $k= \frac{n}{2}$,
 and $m= k, k+1,...$ or $m=-k,-k-1,...$.

The matrix elements of the representations are given in detail in 
\cite{bargmann} \cite{Conrady2}; I quote, for $m'>0,m+m'>0$, the building bloc:
\be
D^k_{m'm}(e^{iL_3\psi}e^{iK_2t}e^{iL_3\varphi})=e^{im'\psi}b^k_{m'm}(t)e^{im\varphi},
\quad b^k_{m'm}(t)=\sqrt{(-1)^{m-m'}}d^k_{m'm}(it),
\nn\ee
and from \cite{bargmann}, the analogue of (\ref{canbasis}):
\begin{align}
\Big(\frac{1}{\sinh t}\frac{\partial}{\partial t}
\sinh t\frac{\partial}{\partial t}&-\frac{1}{\sinh^2 t}
(m^2+m'^2-2mm'\cosh t)\Big)\,b^k_{mm'}(t)=k(k-1)b^k_{mm'}(t)
\label{propwig2}\end{align}
from which we derive the basis functions that diagonalize $(Q,L_3)$:
\be
\psi^{(n\rho)k\pm}_m=C_be^{(i\frac{\rho}{2}-1)b}e^{i(m-\frac{n}{2})\varphi}
b^{\frac{n}{2}}_{m\frac{n}{2}}(t)
\nn\ee
Coherent states for the discrete series can be built as:
\be
v_N=e^{i\bar\varphi L_3}e^{-i\bar tK_1}=
e^{i\sigma_3\frac{\bar\varphi}{2}}e^{\sigma_1\frac{\bar t}{2}}\in SU(1,1)
;\quad  |\psi^{(n\rho)k\pm}_{ N}>:={\cal D}^{k\pm}(v_N)|\psi^{(n\rho)k\pm}_{\pm k}>
\nn\ee
\be
<\psi^{(n\rho)k\pm}_{ N}|(K_1,K_2,L_3)|\psi^{(n\rho)k\pm}_{N}>=
\pm k(\sinh \bar t\sin\bar\varphi,\sinh\bar t\cos\bar\varphi,\cosh\bar t)
\ee
$N$ describes a time-like two-sheeted hyperboloid. 
For the details on the construction of the pseudobasis and of the 
coherent states see \cite{Conrady3}. So far nobody has taken up the challenge of 
using it, work out the asymptotic limit etc.. 

Finally, I want to consider what I regard as the most interesting basis, 
the one based on the abelian 
subgroup of SL(2,$\mathbb{C}$) generated by $(L_3,K_3)$; it has been studied in
	\cite{huszar}\cite{kalnins}\cite{lochlainn}. 
A suitable parametrization is: 
\begin{align}
u_\alpha&=( \tfrac{1}{\sqrt{2}}e^{\frac{c}{2}}e^{\frac{u+i\varphi}{2}},
 \tfrac{1}{\sqrt{2}}e^{\frac{c}{2}}e^{\frac{-u-i\varphi}{2}}),\quad
P^I=e^{c}(\cosh u,\,\cos\varphi,\,\sin\varphi,\,  \sinh u);\cr
L_3&= -i\partial_\varphi+\frac{n}{2},\quad 
L_\pm=e^{\pm i\varphi}(\pm\sinh u\partial_c 
 \mp \cosh u\partial_u +i\sinh u\partial_\varphi +\frac{n}{2}e^{-u}) \cr
K_3&=-i\partial_u,\quad
K_\pm=e^{\pm i\varphi}(i\cosh u\partial_c-i\sinh u\partial_u
\mp \cosh u\partial_\varphi\mp i\frac{n}{2}e^{-u})\cr
\s L^2&=-(\sinh u)^2\partial_u^2-(\cosh u)^2 \partial_\varphi^2
-(\cosh u)^2\partial_u^2 +\sinh 2u\,\partial_u\partial_u+
\partial_u+ne^{-u}\cosh u L_3
\end{align}
The eigenfunctions of  $(L_3,K_3)$ are, with  $m$ (half)integer, 
$\lambda$ real,   $-\infty<\lambda<\infty$:
\begin{align}
\psi^{(n\rho)}_{m\lambda}(c,u,\varphi)&=
\tfrac{\sqrt{2}}{2\pi}\,e^{(i\frac{\rho}{2}-1)c}e^{i(m-\frac{n}{2})\varphi}
e^{i\lambda u}\cr
%
L_3\psi^{(n\rho)}_{m\lambda}&=m\psi^{(n\rho)}_{m\lambda};\quad 
K_3\psi^{(n\rho)}_{m\lambda}=\lambda\psi^{(n\rho)}_{m\lambda}; \quad
<\psi^{(n\rho)}_{m\lambda}|\psi^{(n\rho)}_{m'\lambda'}>=
\delta(\lambda-\lambda')\delta_{mm'}\cr
L_\pm\psi^{(n\rho)}_{m\lambda}&=\psi^{(n\rho)}_{m\pm 1,\lambda}
\big(\cosh u(\frac{n}{2}\pm i\lambda)-\sinh u(m\mp \frac{\rho}{2}\pm 1)\big)\cr
K_\pm\psi^{(n\rho)}_{m\lambda}&=\psi^{(n\rho)}_{m\pm 1,\lambda}
\big(\cosh u(1  \mp im-\frac{\rho}{2}) -\sinh u(\lambda\pm i \frac{n}{2})\big)
\label{theeigenf}\end{align}
 These states have been studied in \cite{smorodinski}\cite{huszar}, and have been used in  
\cite{eugenio} as eigenstates of the 'energy' of a Rindler horizon. To satisfy the simplicity 
constraints we set $\lambda=\gamma m$ for the 3-rd component, and let (\ref{commrel1})
(\ref{commrel2}) take care of the other two. There are no highest weight states, and
therefore no coherent states as we had them before, so the question is whether
we can do without them, and still have states which somehow encode the geometry of the
triangles.

Suppose  we smear these states, using
 $f_\delta(\lambda):=\frac{1}{(2\pi)^{1/4}\delta^{1/2}}e^{-\frac{\lambda^2}{4\delta^2}}$,
and defining 
\be
|m\lambda\,\delta>:=\int d\lambda'f_\delta(\lambda'-\lambda)|(n\rho)m\lambda'>\ 
\leftrightarrow
 \psi^{(n\rho)}_{m\lambda\delta}=\frac{  (2\pi)^{1/4}  \sqrt{\delta}}{\pi}\,
e^{(i\frac{\rho}{2}-1)c}e^{i(m-\frac{n}{2})\varphi}
e^{i\lambda u}e^{-u^2\delta^2}
\label{reg}\ee
With this regulariztion we have that:
\be
<m'\lambda'\,\delta|m\lambda\,\delta>=
\delta_{mm'}e^{-\frac{(\lambda-\lambda')^2}{8\delta^2}},\quad
<m\lambda\,\delta|(L_a+iK_a)|m\lambda\,\delta>=(0,0,m+i\lambda)
\ee
Therefore, if we boost the regularized states with ${\cal D}^{(n\rho)}(g)$ or 
${\cal D}^{(n\rho)}(v)$ we get states such that:
\begin{align}
|\psi^{(n\rho)\frac{n}{2}}_{\s n}>:&={\cal D}^{(n\rho)}(u_{\s n})|m\lambda\,\delta>;\quad
 <\psi^{(n\rho)\frac{n}{2}}_{\s n}|L_a |\psi^{(n\rho)\frac{n}{2}}_{\s n}>=\tfrac{n}{2}\,n_a\cr
|\psi^{(n\rho)k\pm}_{N}>:&={\cal D}^{(n\rho)}(v_N)|m\lambda\,\delta>;\cr
&\quad
<\psi^{(n\rho)k\pm}_{ N}|(K_1,K_2,L_3)|\psi^{(n\rho)k\pm}_{N}>=
\pm k(\sinh t\sin\varphi,\sinh t\cos\varphi,\cosh t)
\nn\end{align}
The probability distribution of the area  in a state of given $(m,\lambda)$ will
be given by the modulus square of the overlaps  of  
$\psi_{\frac{n}{2}}^{(n\rho)\frac{n}{2}}$ or $\psi^{(n\rho)k\pm}_{\pm\frac{n}{2}}$ with
$\psi^{(n\rho)}_{\pm\frac{n}{2}\lambda}$. These have  been calculated\cite{huszar}
\cite{eugenio}\cite{lochlainn}, see Appendix B, and have indeed a maximum at
$\lambda\approx \frac{\rho}{2}$, but further studies are needed to judge whether
these staes provide a viable alternative.

 \section{ A  detailed example of triangulation.}

 A triangulation evolving a pentachoron at t=0 
$(abcde)$ to a later
pentachoron at t=1 $(a'b'c'd'e')$ can be realized connecting with edges all the 
vertices of the first to the vertices of the second, but \underline{omitting} the edges
$(aa'),(bb'),(cc'),(dd'),(ee')$; in this way we realize a division of the 
spacetime $S^3\otimes R$ between t=0 and t=1 in 30 4-simplices. 
This triangulation can be proved to be orientable,
i.e. it can be organized  so that each of the 70 tetrahedron is in two 4-simplices with
opposite orientation, for ex. 
\be
 \begin{matrix}
(abcde'):&  a b c d&a b e' c& a b d e'& a c e' d&b c d e'\\
(abdc'e'):& a b d c'&a b e' d&a b c' e'& a d e' c'& b d c' e'
\end{matrix}
\label{tetras}\ee
The face dual to the triangle $(abe')$ will be made of  the following  four steps:
\be
\begin{matrix}
(abcde') &  & (abdc'e') &  & (abe'c'd') & & (abce'd')  &  & (abcde') \\
 abe'c,abde' &\to&abe'd,abc'e' &\to&abe'c',abd'e'&\to& abe'd',abce' &\to&abe'c,abde'
\end{matrix}\label{face}\ee 

\epsfxsize=200pt\epsfysize=200pt
\begin{figure}[!h]
\begin{minipage}{0.5\linewidth}
\includegraphics[width=\linewidth]{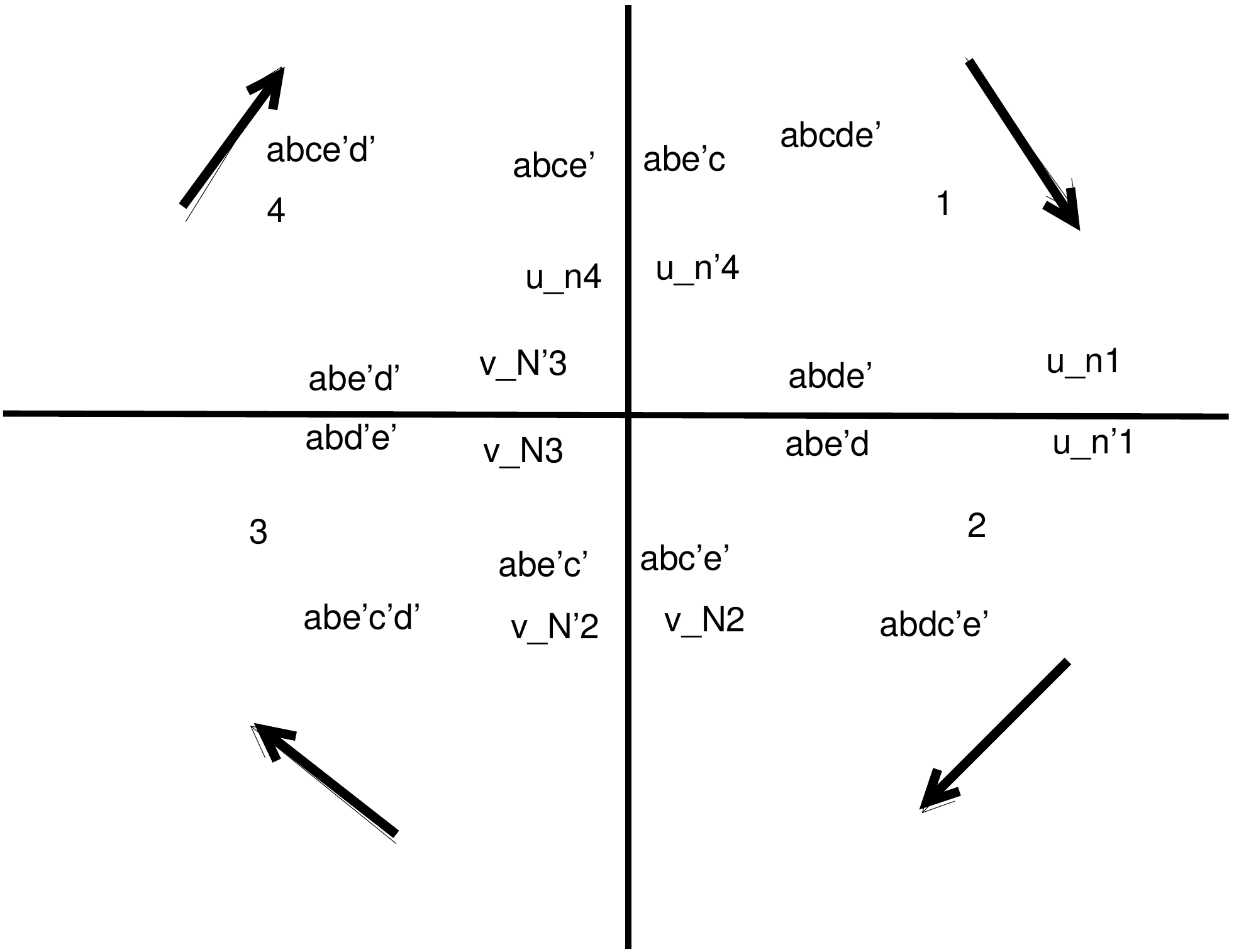}
\end{minipage}\begin{minipage}{0.48\linewidth}
One can see that for ex.  tetrahedron $(abde')=(abcde')\cap (abdc'e')$, 
and that within  $(abcde')$ triangle $t=(abe')=(abe'c)\cap (abde')$. 
4-simplices $(abcde'), (abdc'e')$  have different frames, and
triangle $t=(abe')$ different area tensor, say $S^{IJ}_t,  S'^{IJ}_t$ but
 $S^{IJ}_tS_{t\,IJ}=S'^{IJ}_tS'_{t\,IJ}$.
The two frames are connected by a Lorentz transformation;
 the product of the four  successive  transformations, or
SL(2,C) elements, gives the contribution of  the face.
 Each transformation can be split for ex.
$g_{(abcde')\to (abdc'e')}=g_{(abde')}\,\epsilon\, g_{abe'd}$ (the $\epsilon$ 
accounts for the reversal of orientation), so that the various contributions
can be grouped in a product of `vertex functions', one per 4-simplex.
\end{minipage}\end{figure}

Can we assume that all the tetrahedra involved are space-like?
To get an idea, consider the various 4-simplices involved in our model
triangulation, and to simplify the analysis as much as possible assume that all
edges at time zero have (length)$^2\ l^2 $,  edges at time 1  (length)$^2\ l'^2 $,
 edges between time 0 and time 1  (length)$^2=d^2$. The centre of a bottom 
tetrahedron, say (abcd), is therefore at distance $\sqrt{\frac{3}{8}}l$ from 
any of its vertices, and for the two slices to be separated it must be 
$d^2<\frac{3}{8}l^2$; by the same argument applied to a top tetrahedron
$d^2<\frac{3}{8}l'^2$. To go further we can use Brewin's algorithm\cite{brewin},
or G\"orlich's explicit parametrization\cite{gorlich}.
 I find the following for the normals to the tetrahedra and the (area)$^2$ of 
the triangles :

(4,1): \ $(abcde')$
\begin{align}
V^2_{(abcd)}&=-\half l^6; \   V^2_{(abe'c)}=...=V^2_{(bcde')}=
\tfrac{3}{4}l^4(\tfrac{1}{3}l^2-d^2);\cr
S^2_{dbc}&=S^2_{cad}=S^2_{bac}=S^2_{abd}=\frac{3}{8}l^4,\quad
S^2_{dce}=...=S^2_{eac}=\frac{l^2}{2}(d^2-\frac{1}{4}l^2).
\nn\end{align}
(here $S^2:=S^{IJ}S_{IJ}$).
As we have seen, if $d^2=\tfrac{3}{8}l^2$ the two hypersurfaces are stuck together;
 they come apart as $d^2$ decreases, and all the tetrahedra  are space-like, but 
four of them become light-like when $d^2=\tfrac{1}{3}l^2$, and time-like for 
still smaller $d^2$; six (area)$^2$ vanish at  $ d^2=\tfrac{1}{4}l^2$
and become negative for smaller $d^2$. So  in the  'good'  interval 
$\tfrac{3}{8}l^2>d^2>\tfrac{1}{4}l^2$ all $S^2$ are positive, and for 
$\tfrac{3}{8}l^2>d^2>\tfrac{1}{3}l^2$ all tetrahedra are space-like.
Of course the same analysis applies to the (1,4) four simplices, 
with $l\leftrightarrow l'$.

(3,2): \ $(abdc'e')$
\begin{align}
V^2_{(abdc')}&=V^2_{(abe'd)}=\tfrac{3}{4}l^4(\tfrac{1}{3}l^2-d^2);\quad
V^2_{(abc'e')}=V^2_{(ade'c')}=V^2_{(bdc'e')}=l^2{l'}^2(\quart l^2+\quart l'^2-d^2)\cr
S^2_{abd}&=\frac{3}{8}l^2;\quad
S^2_{ac'e'}= S^2_{dc'e'}=S^2_{c'be'}=\frac{l'^2}{2}(d^2-\frac{1}{4}l'^2);\quad 
S^2_{abe'}=...=S^2_{dae'}=\frac{l^2}{2}(d^2-\frac{1}{4}l^2)
\nn\end{align}
We have to look at this case  together with the previous ones and the (2,3) case;
 given that  $d^2$ is such that   $d^2<\frac{3}{8}l^2,\ d^2<\frac{3}{8}l'^2$, it must
be $d^2<\frac{3}{16}(l^2+l'^2)$; but then tetrahedra (abc'e'), (bdc'e'), (ade'c') cannot 
be space-like, because that would require $d^2>\frac{1}{4}(l^2+l'^2)$.
We cannot therefore have this triangulation made exclusively of
space-like tetrahedra.  On the other hand the (area)$^2$ of the triangles stay
positive provided $d^2>\quart l^2,\ d^2>\quart l'^2$. Notice the role that the 
various tetrahedra play in the face dual to (abe') in (\ref{face}).

It may be that our simplifying  assumptions are too strong, and that one can 
find a choice of lenghts which makes all tetrahedra space-like. I find it
encouraging that one can at least assume that all (area)$^2$ are positive,
which is what matters.
It may also be that larger triangulations do not have such tight constraint;
this would require some experimenting  that I have not attempted.

 For the face dual to the triangle $(abe')$ one would therefore
find, on the assumption that tetrahedra 
$(abce'), (abde')$ are space-like, tetrahedra $(abc'e'), (abd'e')$ time-like,
simplifying as much as possible the notation:
\begin{align}
f_{(abe')}&=c
\cdot\psi_{j,\gamma j}^\dagger u_{\s n'_4}^{-1}g'^{-1}_4 \epsilon g_1u_{\s n_1}\psi_{j,\gamma j}
\psi_{j,\gamma j}^\dagger u_{\s n'_1}^{-1}g'^{-1}_1\epsilon g_2v_{N_2}\psi_{j,\gamma j}
\psi_{j,\gamma j}^\dagger v_{N'_2}^{-1}g'^{-1}_2\epsilon g_3v_{N_3}\psi_{j,\gamma j}\cdot\cr
& \qquad\qquad\qquad\qquad \qquad\qquad  \qquad\qquad          \qquad\cdot
\psi_{j,\gamma j}^\dagger v_{N'_3}^{-1}g'^{-1}_3\epsilon g_4u_{\s n_4}\psi_{j,\gamma j}
\nn\end{align}
Here I have written $\psi_{j,\gamma j}$ for 
$\psi^{(n_f\rho_f)}_{\tfrac{n}{2},\gamma\tfrac{n}{2}}$, and the representative
of the  SL(2,$\mathbb{C}$) elements ${\cal D}^{(n_f\rho_f)}(g)$,... simply as $g$,... .

The `vertex function' of a 4-simplex, say $(abcde')$ is the product of the factors
 corresponding to its five tetrahedra, integrated over the group variables, i.e. 
 the product of matrix elements between  data for the  10 triangles.
Because of the overall SL(2,$\mathbb{C}$) invariance, one of the
group integration can be suppressed, and the final expression is finite.
To calculate the partition function for the triangulation the product of the
vertex functions should finally be summed over the $n_f$.

\bigskip\bigskip
\noindent{\bf Conclusions}

Spinors are a valuable tool to formulate the spin-foam theory, and they might
allow the construction of  calculable models. They give a considerable 
simplification of the various technical problems, which have so far hidden the basic 
simplicity of the model; hopefully, they might be instrumental in going beyond the
 simple picture of space-time at the basis of it. The use of the spinor basis for the
quantum theory needs, and perhaps deserves further study.  

\bigskip
\noindent{\bf Acnowledgment}

I am very grateful to Dimitri Marinelli for the many long discussions we have 
had over the years, among other things on the subject of this work, and for help
with some calculations, and to 
Eugenio Bianchi for telling me about \cite{huszar} and  sending me a copy of it.
\bigskip
\appendix\noindent{\bf Appendix A: 4-vectors, tensors, spinors.}

I collect here the various definitions and identities used in the text. 

Dealing with spinors, I   use repeatedly the following  identities for  
the Pauli matrices $\sigma_a$ and the matrix  $\epsilon=i\sigma_2$: 
\be
\epsilon\sigma_a\epsilon^{-1}=-\sigma_a^T;\quad
\sigma_{a\alpha\beta}\sigma_{a\gamma\delta}=\delta_{\alpha\beta}\delta_{\gamma\delta} -
2\epsilon_{\alpha\gamma}\epsilon_{\beta\delta};\quad
t_\alpha\epsilon_{\alpha\beta}u_\beta \,v_\gamma+
u_\alpha\epsilon_{\alpha\beta}v_\beta \,t_\gamma+
v_\alpha\epsilon_{\alpha\beta}t_\beta \,u_\gamma=0.
 \label{spinorida}\ee
For the 4-d sigma matrices we have:
\begin{align} 
\sigma^I&=(1,\sigma_a),\quad 
\tilde\sigma^I=(-1,\sigma_a)=-\epsilon\,\bar\sigma^{I}\, \epsilon^{-1};\cr
\eta^{IJ}&=\half \tr(\sigma^I\tilde\sigma^J);\quad 
\sigma_I\tilde\sigma_J+\sigma_J\tilde\sigma_I=2\eta_{IJ};\quad
\half\epsilon^{IJKL}\sigma_K\tilde\sigma_L=i\sigma^{[I}\tilde\sigma^{J]}.
\label{trfvectors}
\end{align}
The last equation can be used to project the  selfdual/antiselfdual 
 or the left/right components of an antisymmetric tensor:
\begin{align}
\sigma_I\tilde\sigma_JJ^{IJ}&=\sigma_I\tilde\sigma_JJ^{IJ}_+=
4i\sigma_aJ^L_a
,\quad\tilde\sigma_I\sigma_JJ^{IJ}=\tilde\sigma_I\sigma_JJ^{IJ}_-=4i\sigma_aJ^R_a
\label{trftensors}\end{align}
where $J_\pm^{IJ}=\half(J^{IJ}\mp i\,   { ^*J}^{IJ})$. 
 I have used the identities, valid for  an arbitrary 4-vector $V^I$:
\be
V_M\tilde\sigma_IJ^{MI}_{+}=i\tilde\sigma_IV^I\,\sigma_a J^L_a;\qquad
V_M\tilde\sigma_IJ^{MI}_{-}=-i\sigma_a J^R_a \,\tilde\sigma_IV^I.
\label{spinoridb}
\ee
Proof: from the definitions we have 
$J^{ab}_+=\epsilon_{abc}J^L_c,\ J^{0a}_+=-iJ^L_a$; then:
\begin{align}
V_M\tilde\sigma_IJ^{MI}_+&=V_aJ^{a0}_+ +V_0\sigma_aJ^{0a}_+ +V_b\sigma_cJ^{bc}_+=
iV_aJ^L_a-iV_0\sigma_aJ^L_a-V_b\sigma_c\epsilon_{cba}J^L_a=\nn\\
&=iV_aJ^L_a-iV_0\sigma_aJ^L_a+iV_b(\sigma_b\sigma_a-\delta_{ab})J^L_a=
i(V^0+\sigma_bV^b)\sigma_aJ^L_a . 
\nn\end{align}

The correspondence between spinor and vector representations of  SL(2,$\mathbb C$)
 is fixed by: 
\be
g\sigma_Ig^\dagger= \sigma_J\Lambda^J_{\ I};\quad 
g^{\dagger -1}\tilde\sigma_Ig^{-1}=\tilde \sigma_J\Lambda^J_{\ I};\quad
g\sigma_ag^{-1}=\sigma_bO_{ba},
\label{sl2ctrf1}\ee
where the orthogonal matrix $O_{a'a}$ is the representative of  $g$ in the(1,0) 
representation.
 The action of  SL(2,$\mathbb C$) on 4-vectors and on the basic spinors is given by:
\be
T_g\cdot u= ug; \quad T_g\cdot t= t g^{\dagger -1};\quad 
T_g\cdot X^I=X^{I'}\Lambda_{I'}^{\ I};\quad 
T_g\cdot(u\sigma_I\bar u)= (u\sigma_{I'}\bar u)\,\Lambda_{I'}^{\ I}
 \label{sl2ctrf2}\ee
 These transformations are implemented through the  SL(2,$\mathbb C$)
invariant Poisson brackets \\$\{t_\alpha,\bar u_\beta\}=-i\delta_{\alpha\beta};\quad
\{u_\beta,\bar t_\alpha\}=-i\delta_{\alpha\beta}$, or
\be
\{f,h\}=i\frac{\partial f}{\partial \bar t_\alpha}\frac{\partial h}{\partial  u_\alpha}
-i\frac{\partial f}{\partial  u_\alpha}\frac{\partial h}{\partial \bar t_\alpha}
-i\frac{\partial f}{\partial  t_\alpha}\frac{\partial h}{\partial  \bar u_\alpha}
+i\frac{\partial f}{\partial \bar u_\alpha}\frac{\partial h}{\partial  t_\alpha},
\label{pb1}\ee
and generators:
\be
J^L_a=-\half u\sigma_a\bar t,\quad J^R_a=-\half t\sigma_a\bar u;\quad
\{J^{L,R}_a,J^{L,R}_b\}=\epsilon_{abc}J^{L,R}_c,\quad\{J^L_a,J^R_b\}=0.
\label{generators}\ee
One finds:
\be
\{u_\alpha,J^L_a\}=\ihalf u_\beta\sigma_{a\beta\alpha};\quad
\{f(u,\bar u),J^L_a\}=\ihalf u_\beta\sigma_{a\beta\alpha}\frac{\partial f}{\partial u_\alpha};
\quad \{f(u,\bar u),J^R_a\}=\ihalf  \frac{\partial f}{\partial\bar u_\alpha}
\sigma_{a\alpha\beta}\bar u_\beta
\label{pb2}\ee
and therefore, if $g=1+i\epsilon_aL_a+i\eta_aK_a=1 +(i\epsilon_a +\eta_a)J^L_a+
(i\epsilon_a -\eta_a)J^R_a\simeq 1+\frac{i}{2}(\epsilon_a\sigma_a+i\eta_a\sigma_a)$
\be
T_{(\epsilon\eta)}\cdot f(u,\bar u)=
\ihalf(i\epsilon_a+\eta_a)u_\beta\sigma_{a\beta\alpha}\frac{\partial f}{\partial u_\alpha} +
\ihalf(i\epsilon_a-\eta_a)\frac{\partial f}{\partial \bar u_\alpha}\sigma_{a\alpha\beta}
\label{pb3}\ee
To go to finite group elements:
\be
e^{ \{ \, .\,, v_aJ^L_a+w_aJ^R_a \} }f(u,\bar u)= f(ue^{\ihalf v_a\sigma_a},
\bar ue^{\ihalf w_a\sigma_a}).
\label{pb4}\ee

\bigskip
\appendix\noindent{\bf Appendix B: The Overlap}

The overlap of  
$\psi_{\frac{n}{2}}^{(n\rho)\frac{n}{2}}$, $\psi^{(n\rho)k\pm}_{\pm\frac{n}{2}}$ with
$\psi^{(n\rho)}_{\pm\frac{n}{2}\lambda}$ can be calculated 
considering the states of the three types considered, with $\frac{n}{2}=j=k,\ m=\pm n$;
the expressions of $d^{\frac{n}{2}}_{\frac{n}{2}\frac{n}{2}}(\theta),
b^{\frac{n}{2}}_{\pm\frac{n}{2}\pm\frac{n}{2}}(t)$ are very simple. Listing them, with
the parametrizations of $P^I$:
\begin{align}
  \psi_{\frac{n}{2}}^{(n\rho)\frac{n}{2}}&=
C_ae^{(i\frac{\rho}{2} -1)a}d^{\frac{n}{2}}_{\frac{n}{2}\frac{n}{2}}(\theta)=
C_ae^{(i\frac{\rho}{2} -1)a}\Big(\frac{1+\cos\theta}{2}\Big)^{\frac{n}{2}};
\quad P^I= e^a(1,\sin\theta\cos\varphi,\sin\theta\sin\varphi,\cos\theta)\cr
\psi^{(n\rho)k\pm}_{\pm\frac{n}{2}}&=C_be^{(i\frac{\rho}{2}-1)b}
b^{\frac{n}{2}}_{\pm\frac{n}{2}\pm\frac{n}{2}}(t)=
C_be^{(i\frac{\rho}{2}-1)b}\Big(\frac{2}{1+\cosh t}\Big)^{\frac{n}{2}};
 \ P^I=e^b(\cosh t,\sinh t\cos\varphi,\sinh t\sin\varphi,1)\cr
\psi^{(n\rho)}_{\pm\frac{n}{2}\lambda}&=
C_c\,e^{(i\frac{\rho}{2}-1)c} e^{i\lambda u};
\qquad \qquad P^I=e^c(\cosh u,\cos\varphi,\sin\varphi,\sinh u)\cr
&\frac{d^3P}{2 P(2\pi)^3}=\frac{e^{2a}da\sin\theta d\theta d\varphi}{2(2\pi)^3}=
 \frac{ e^{2b}\sinh t\, dt\,db\, d\varphi}{2(2\pi)^3}=
\frac{e^{2c}dc\, du\, d\varphi}{2(2\pi)^3}. \label{listbasis}
\end{align}
However, as explained in the text, one omits the integration in $da$ or $db$ or $dc$.
Equating the expressions of $P^I$: 
\be
e^a=e^b\cosh t=e^c\cosh u; \quad \cos\theta=\frac{\sinh u}{\cosh u}; \quad 
 \cosh t=\frac{\cosh u}{\sinh u};\quad  
e^b=e^c\sinh u  
\nn\ee
Then I find, in the two cases:
\begin{align}
<\psi^{(n\rho)}_{\frac{n}{2}\lambda}|  \psi_{\frac{n}{2}}^{(n\rho)\frac{n}{2}}>&=
\int\frac{e^{2c} du\, 2\pi}{2(2\pi)^3} C_cC_a\,e^{(-i\frac{\rho'}{2}-1)c}
e^{-i\lambda u}e^{(i\frac{\rho}{2} -1)a}\Big(\frac{1+\cos\theta}{2}\Big)^{\frac{n}{2}}=\cr
&=\frac{C_cC_a}{8\pi^2\, 2^{n/2} }\int du
 \frac{e^{(-i\lambda+\frac{n}{2}) u}}{(\cosh u)^{1+\frac{n}{2}-i\frac{\rho}{2}}}=\cr
&= \frac{C_cC_a}{8\pi^2\, 2^{(i\frac{\rho}{2})} }  
\frac{\Gamma(\tfrac{1}{2}+\tfrac{i}{2}(\lambda-\frac{\rho}{2}))
\Gamma(\tfrac{1}{2}+\tfrac{n}{2}-\tfrac{i}{2}(\lambda+\frac{\rho}{2}))}
{\Gamma(1+\frac{n}{2}-i\frac{\rho}{2})}.\label{case1}\\
<\psi^{(n\rho)}_{\pm\frac{n}{2}\lambda}|\psi^{(n\rho)k\pm}_{\pm\frac{n}{2}}>&=
\int\frac{e^{2c} du\, 2\pi}{2(2\pi)^3}
C_cC_b\,e^{(-i\frac{\rho'}{2}-1)c}e^{-i\lambda u}
e^{(i\frac{\rho}{2} -1)b}\Big(\frac{2}{1+\cosh t}\Big)^{\frac{n}{2}}=\cr
&=\frac{C_cC_b  }{8\pi^2 }2^{n/2} \int du\,
e^{-(i\lambda+\frac{n}{2})u} (\sinh u)^{\frac{n}{2}+i\frac{\rho}{2}-1}=\cr
&=\frac{C_cC_b  }{8\pi^2 \,2^{i\frac{\rho}{2}}}
\frac{\Gamma(\frac{1}{2}+\frac{i}{2}(\lambda-\frac{\rho}{2}))
           \Gamma(\frac{n}{2}+i\frac{\rho}{2})}
{\Gamma(\frac{1}{2}+\frac{n}{2}+\frac{i}{2}(\lambda+\frac{\rho}{2}))}.
\label{case2}\end{align}
From these expressions we get that for $k\ge 1$, up to factors independent on$\lambda$:
\begin{align}
|<\psi^{(n\rho)}_{\frac{n}{2}\lambda}|  \psi_{\frac{n}{2}}^{(n\rho)\frac{n}{2}}>|^2
&\sim\frac{1}{\cosh\pi\lambda+\cosh\pi\frac{\rho}{2}}
\prod_{j=1}^k\frac{(2j-1)^2+(\lambda+\frac{\rho}{2})^2}{4},\quad n=2k;\cr
&\sim
\frac{(\lambda-\frac{\rho}{2}) }{\sinh\pi\lambda-\sinh\pi\frac{\rho}{2}}
\prod_{j=1}^k(j^2+\frac{(\lambda+\frac{\rho}{2})^2}{4}),\quad\quad n=2k+1.\cr
|<\psi^{(n\rho)}_{\pm\frac{n}{2}\lambda}|\psi^{(n\rho)k\pm}_{\pm\frac{n}{2}}>|^2&
\sim\frac{\cosh\frac{\pi}{2}(\lambda+\frac{\rho}{2})}{\cosh\frac{\pi}{2}(\lambda-\frac{\rho}{2})}
 \prod_{j=1}^k\frac{4}{(2j-1)^2+(\lambda+\frac{\rho}{2})^2},\quad\quad n=2k,\cr
&\sim \frac{\sinh\frac{\pi}{2}(\lambda+\frac{\rho}{2})}
{\frac{\pi}{2}(\lambda+\frac{\rho}{2})\cosh\frac{\pi}{2}(\lambda-\frac{\rho}{2})}
\prod_{j=1}^k\frac{1}{j^2+\frac{(\lambda+\frac{\rho}{2})^2}{4}   },\quad n=2k+1.
\end{align}
(for $k=0$, i.e. for $n=0$ or $n=\tfrac{1}{2}$, same expressions with no products).

For $k=1, n=2$ I find with the first of these expressions  
$<\lambda>=\frac{\rho}{4}$ and \\
$<(\lambda-<\lambda>)^2>=\tfrac{3}{5}+\tfrac{3}{20}(\frac{\rho}{2})^2$, and I presume
that the other expressions would give similar results. 
I do not know of a good approximation to investigate the behaviour 
 for large $n$.

\begingroup\raggedright 
\endgroup

\end{document}